\documentclass[sigconf]{acmart} 
\usepackage{graphicx}
\usepackage{subfigure}
\usepackage[utf8x]{inputenc}
\AtBeginDocument{%
  \providecommand\BibTeX{{%
    \normalfont B\kern-0.5em{\scshape i\kern-0.25em b}\kern-0.8em\TeX}}}

\acmDOI{}

\acmBooktitle{} 
\acmPrice{}
\acmISBN{}

\begin{document}
	
	\title{Interpretable Melody Generation from Lyrics with Discrete-Valued Adversarial Training}
	
	\author{Wei Duan, Zhe Zhang, Yi Yu, Keizo Oyama}
	\email{{weiduan,zhe,yiyu,oyama}@nii.ac.jp}
	\affiliation{\institution{National Institute of Informatics, SOKENDAI}
		\state{Tokyo}
		\country{Japan}}

	\renewcommand{\shortauthors}{Wei Duan and Zhe Zhang, et al.}
	
	\begin{abstract}
	Generating melody from lyrics is an interesting yet challenging task in the area of artificial intelligence and music. However, the difficulty of keeping the consistency between input lyrics and generated melody limits the generation quality of previous works. In our proposal, we demonstrate our proposed interpretable lyrics-to-melody generation system which can interact with users to understand the generation process and recreate the desired songs. To improve the reliability of melody generation that matches lyrics, mutual information is exploited to strengthen the consistency between lyrics and generated melodies. Gumbel-Softmax is exploited to solve the non-differentiability problem of generating discrete music attributes by Generative Adversarial Networks (GANs). Moreover, the predicted probabilities output by the generator is utilized to recommend music attributes. Interacting with our lyrics-to-melody generation system, users can listen to the generated AI song as well as recreate a new song by selecting from recommended music attributes.
	\end{abstract}
	

	\maketitle
	
	\section{Background}
	
	With the increasing of availability of various music data, cross-modal music retrieval and generation tasks \cite{LSTM-GAN,cross} are drawing more and more attention in the artificial intelligence, music, and multimedia research community. However, melody generation from lyrics is still a less explored task due to the difficulty of learning the correlation relationship between lyrics and melody. The earliest research of music generation emerged in the middle of 1950s by exploring the mathematical relationships between musical features using computational techniques such as Markov models \cite{early1950}. With the development of available lyrics-melody dataset and deep learning techniques, more and more AI techniques have been exploited in music generation \cite{cond-12,cond-13,cond-15,2017Aly,cond-09, Jukebox, songmass, telemelody, Xiaobing, MusicTransformer, pop, tbc}. However, how to generate realistic melody from lyrics that preserves the meaningful correlation between lyrics and melody remains a challenge.
	
	In this demonstration paper, we develop a lyrics-to-melody generation system, which has the capability of predicting a melody when the lyrics is provided by human users. Human users are also able to interact with the system to recreate new songs. To the best of our knowledge, this is the first interactive lyrics-conditioned neural melody generation system. The lyrics are tokenized into word level and syllable level sequences. Then the token sequences are both encoded by their corresponding encoder network based on Skip-gram models \cite{Skip}. Taking the advantages of Generative Adversarial Networks (GANs) \cite{GAN,2018GAN}, the lyrics-conditioned LSTM-based generator and discriminator network decode the inner representations to generate the corresponding melody attributes, namely pitch, duration, and rest. Moreover, mutual information is exploited in the model to help the model avoid losing the meaningful information in the lyrics. Gumbel-Softmax is utilized to solve the non-differentiability problem of generating discrete melody attributes by GANs. Finally, the  melody is generated in the MIDI format with readable sheet music, while the candidates of melody attributes are offered by our system for human users to select and recompose new songs.
	
	\section{Lyrics-conditioned melody generation system}
	
\begin{figure}[h]
		\vspace{-0.2cm}
		\centering
		\includegraphics[width=\linewidth]{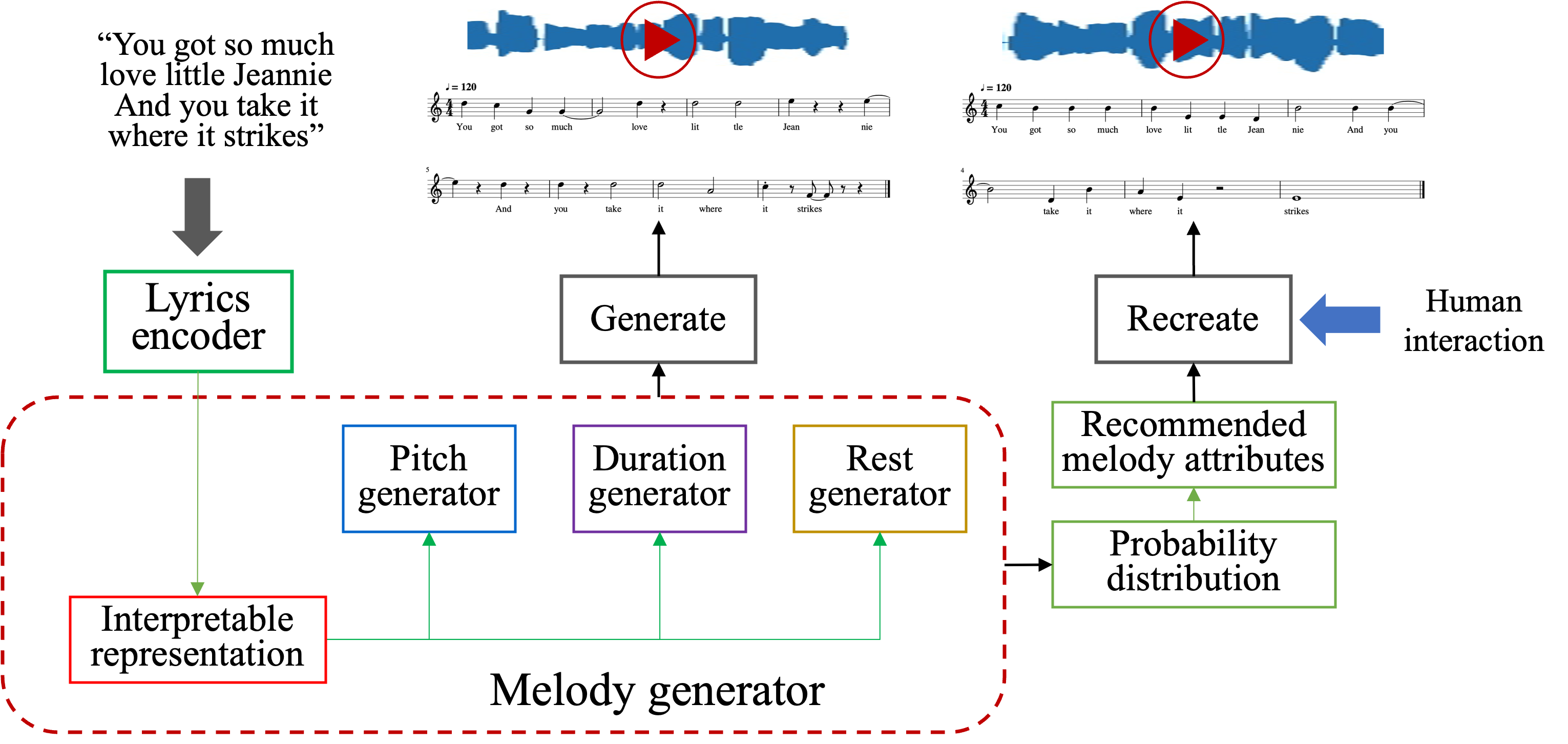}
		\caption{Interpretable lyrics-conditioned melody generation and re-creation system.}
		\label{fig:architecure}
	\end{figure}

	The proposed AI lyrics-to-melody generation system mainly consists of three parts as shown in Fig. \ref{fig:architecure}: lyrics encoder, lyrics-conditioned melody generation with GAN training, and an interactive interface for human users to generate AI songs and further select AI-recommended melody attributes based on the model predicted probabilities to recreate new songs as well as sheet music.

	\subsection{Lyrics encoder}
	We use a large paired lyrics-melody music dataset proposed by \cite{LSTM-GAN} to train our model. On this basis, syllable-level and word-level Skip-gram models are respectively trained over the entire lyrics dataset to obtain a different-level semantic representation of the lyrics. In particular, the lyrics are embedded into 10-dimensional syllable-level vectors and 10-dimensional word-level embedding vectors respectively, which are concatenated together as a more meaningful semantic lyrics embedding vector.
	
	\subsection{Lyrics-to-melody generation}
	
	The melody generation network takes the encoded lyrics embedding vectors as input and generates sequences of melody attributes as output including pitch, duration, and rest. 
	
	Based on the conditional GAN architecture, lyrics embedding vectors as conditions are introduced into both the generator and discriminator in the model training process. The generator and discriminator are based on LSTM networks \cite{tbc_lstm}.
	
	Using GAN to generate discrete-valued melody sequences would face a critical problem called non-differentiability. The generator cannot get the gradient of the discriminator during the training process, thus its parameters cannot be updated. To address this problem, our model introduces the Gumbel-Softmax to derive the differential approximation of discrete music attributes in the GAN training. Finally, with the help of the Gumbel-Softmax trick, by optimizing the relativistic standard GAN (RSGAN) loss function \cite{GANloss}, the proposed model can generate realistic melody that matches the input lyrics after training. 
	
	\subsection{Mutual information constraints}
	
	In our architecture, the mutual information constraint \cite{Info} is applied to reduce the loss of information and keep the consistency between the lyrics and the generated melodies while improving generation quality.
	
	In this paper, our work takes the mutual information constraint to make interpretable vectors $M(x)$ and the lyrics embedding $x$ as similar as possible to improve the interpretability and consistency. However, it is hard for us to directly calculate the mutual information $I (x; M (x))$. Therefore we exploit the posterior approximation layer $Q$ to compute the lower bound of mutual information $I (x; M (x))$. After training the model with the learning objective of mutual information, the information of syllables in the training process is preserved as much as possible, which ensures the strong content consistency between input lyrics and generated melodies.
	
	\section{Experiment and Demonstration}
\begin{figure}[h]
		\centering
		\includegraphics[scale=0.33]{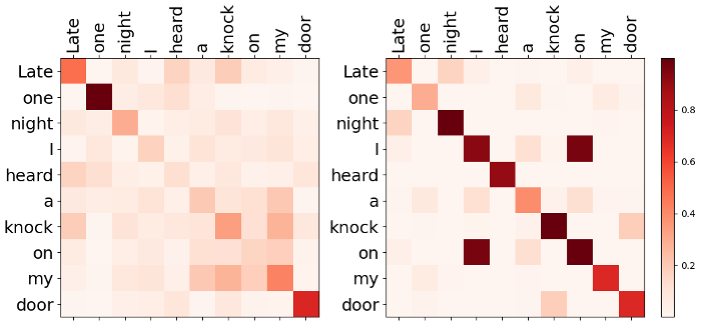}
		\caption{Heatmap between syllables.}
		\label{fig:heatmap}
	\end{figure}

	In this section, we exploit syllable embeddings and interpretable representations to plot the heatmap which shows the interpretability between embedding features of syllables. In the Fig. \ref{fig:heatmap}, the left and right figures respectively visualize weights of syllable embeddings without and with mutual information constraint. In these figures, the block indicates there is a correlation between corresponding syllables. The darker the color is, the more relative there is. Therefore, compared with left figure, the right figure shows that the mutual information makes syllable embedding more accurate and further improves the ability to learn the correlation between syllables and melody attributes. It demonstrates that the mutual information can reduce the information loss of syllables.
	
\begin{figure}[htbp]
		\centering
		\subfigure[]{
			\begin{minipage}[t]{0.5 \linewidth}
				\centering
				\includegraphics[scale=0.16]{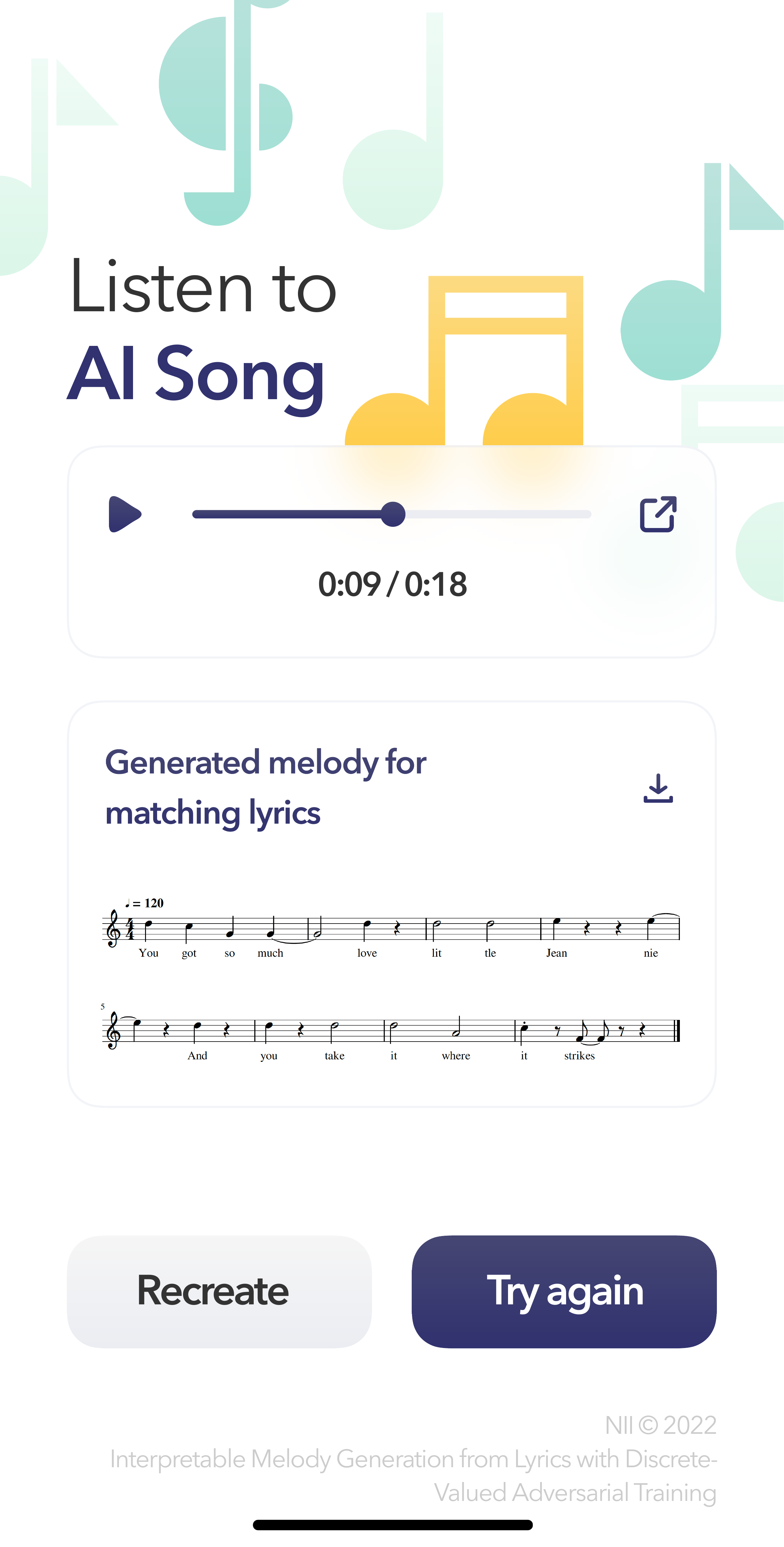}
			\end{minipage}%
		}%
		\subfigure[]{
			\begin{minipage}[t]{0.5 \linewidth}
				\centering
				\includegraphics[scale=0.16]{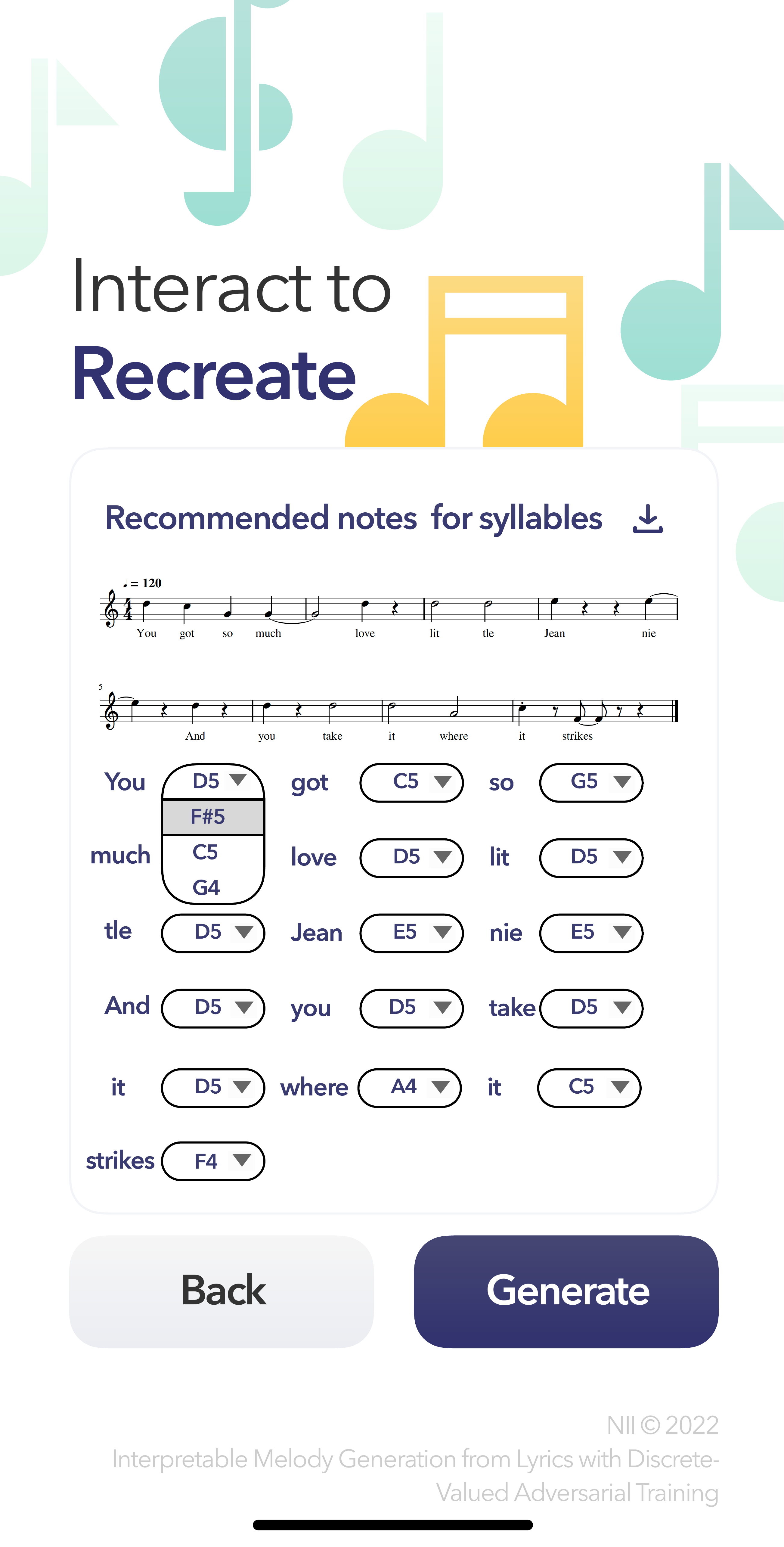}
			\end{minipage}
		}%
		\centering
		\caption{Screenshots of demo.}
		\label{fig:demo}
		\vspace{-0.5cm}
	\end{figure}

	In addition, Fig. \ref{fig:demo} demonstrates the GUI of our interactive lyrics-conditioned melody generation and re-creation system. When given lyrics, the system predicts melody, and the alignment between syllables and melody attributes, and on this basis produces sheet music. After viewing and listening to the AI-composed song, human users can also recreate a piece of melody by choosing from recommended melody attributes offered by our system. The lyrics and recomposed melody are further synthesized into music audio and sheet music for users. By involving human users in the generation process, the system demonstrates the human-centered AI and the concept of Human-in-the-Loop. 
	
\begin{thebibliography}{99}  
		\bibitem{2017Aly}Margareta Ackerman and David Loker, "Algorithmic Songwriting with ALYSIA," \emph{6th International Conference on Computational Intelligence in Music, Sound, Art and Design (EvoMUSART)} , pp.1-16, 2017.
		
		\bibitem{Info}Xi Chen, Yan Duan, Rein Houthooft, John Schulman, Ilya Sutskever, and Pieter Abbeel, "InfoGAN: Interpretable Representation Learning by Information Maximizing Generative Adversarial Nets," \emph{29th International Conference on Neural Information Processing Systems (NIPS)}, pp.2172-2180, 2016.
		
		\bibitem{Jukebox}Prafulla Dhariwal, Heewoo Jun, Christine Payne, Jong Wook Kim, Alec Radford, and Ilya Sutskever, "Jukebox: A Generative Model for Music," \emph{CoRR} abs/2005.00341, 2020.
		
		\bibitem{2018GAN}Hao-Wen Dong, Wen-Yi Hsiao, Li-Chia Yang, and Yi-Hsuan Yang, "MuseGAN: Multi-track Sequential Generative Adversarial Networks for Symbolic Music Generation and Accompaniment," \emph{32nd AAAI Conference on Artificial Intelligence}, pp.34-41, 2018.
		
		\bibitem{GAN}Ian J. Goodfellow, Jean Pouget-Abadie, Mehdi Mirza, Bing Xu, David Warde-Farley, Sherjil Ozair, Aaron C. Courville, and Yoshua Bengio, "Generative Adversarial Networks," \emph{CoRR} abs/1406.2661, 2014.
		
		\bibitem{tbc_lstm}Sepp Hochreiter and J{\"{u}}rgen Schmidhuber, "Long Short-Term Memory," \emph{Neural Computation}, Vol. 9, No. 8, pp.1735-1780, 1997.
		
		\bibitem{MusicTransformer}Cheng-Zhi Anna Huang, Ashish Vaswani, Jakob Uszkoreit, Ian Simon, Curtis Hawthorne, Noam Shazeer, Andrew M. Dai, Matthew D. Hoffman, Monica Dinculescu, and Douglas Eck, "Music Transformer: Generating Music with Long-Term Structure," \emph{7th International Conference on Learning Representations (ICLR)}, 2019.
		
		\bibitem{GANloss}Alexia Jolicoeur-Martineau, "The relativistic discriminator: a key element missing from standard GAN," \emph{7th International Conference on Learning Representations (ICLR)}, 2019.
		
		\bibitem{telemelody}Zeqian Ju, Peiling Lu, Xu Tan, Rui Wang, Chen Zhang, Songruoyao Wu, Kejun Zhang, Xiangyang Li, Tao Qin, and Tie-Yan Liu, "TeleMelody: Lyric-to-Melody Generation with a Template-Based Two-Stage Method," \emph{CoRR} abs/2109.09617, 2021.
		
		\bibitem{Skip}Tom{\'{a}}s Mikolov, Kai Chen, Greg Corrado, and Jeffrey Dean, "Efficient Estimation of Word Representations in Vector Space," \emph{1st International Conference on Learning Representations (ICLR)}, 2013.
		
		\bibitem{cond-12}Kristine Monteith, Tony R. Martinez, and Dan Ventura, "Automatic Generation of Melodic Accompaniments for Lyrics," \emph{3rd International Conference on Computational Creativity (ICCC)}, pp.87-94, 2012.
		
		\bibitem{cond-09}Eric Nichols, "Lyric-Based Rhythm Suggestion," \emph{35th 2009 International Conference on International Computer Music Conference (ICMC)}, 2009.
		
		\bibitem{pop}Yi Ren, Jinzheng He, Xu Tan, Tao Qin, Zhou Zhao, and Tie-Yan Liu, "Pop-MAG: Pop Music Accompaniment Generation," \emph{28th ACM International Conference on Multimedia (ACM MM)}, pp.1198-1206, 2020.
		
		\bibitem{cond-15}Marco Scirea, Gabriella A. B. Barros, Noor Shaker, and Julian Togelius, "SMUG: Scientific Music Generator," \emph{6th International Conference on Computational Creativity (ICCC)}, pp.204-211, 2015.
		
		\bibitem{songmass}Zhonghao Sheng, Kaitao Song, Xu Tan, Yi Ren, Wei Ye, Shikun Zhang, and Tao Qin. 2021, "SongMASS: Automatic SongWriting with Pre-training and Alignment Constraint," \emph{35th AAAI Conference on Artificial Intelligence}, pp.13798-13805, 2021
		
		\bibitem{tbc}Abhishek Srivastava, Wei Duan, Rajiv Ratn Shah, Jianming Wu, Suhua Tang, Wei Li, and Yi Yu, "Melody Generation from Lyrics Using Three Branch Conditional LSTM-GAN," \emph{28th International Conference on MultiMedia Modeling (MMM)} pp.569-581, 2022.
		
		\bibitem{cond-13}Jukka M. Toivanen, Hannu Toivonen, and Alessandro Valitutti, "Automatical Composition of Lyrical Songs," \emph{4th International Conference on Computational Creativity (ICCC)}, pp.87-91, 2013.
		
		\bibitem{early1950}Peter Westergaard, "Experimental Music. Composition with an Electronic Computer," 1959.
		
		\bibitem{LSTM-GAN}Yi Yu, Abhishek Srivastava, and Simon Canales, "Conditional LSTM-GAN for Melody Generation from Lyrics," \emph{ACM Transaction on Multimedia Computing Communication and Applications (TOMCCAP)}, Vol. 17, No. 1, article 35, pp.1-20 2021.
		
		\bibitem{cross}Yi Yu, Suhua Tang, Francisco Raposo, and Lei Chen, "Cross-modal correlation learning for audio and lyrics in music retrieval," \emph{ACM Transaction on Multimedia Computing Communication and Applications (TOMCCAP)}, Vol.15, No. 1, article 20, pp.1-16, 2019.
		
		\bibitem{Xiaobing}Hongyuan Zhu, Qi Liu, Nicholas Jing Yuan, Chuan Qin, Jiawei Li, Kun Zhang, Guang Zhou, Furu Wei, Yuanchun Xu, and Enhong Chen, "XiaoIce Band: A Melody and Arrangement Generation Framework for Pop Music," \emph{24th ACM SIGKDD International Conference on Knowledge Discovery \& Data Mining, (KDD)},  pp.2837-2846, 2018. 
		
	\end{thebibliography}
	
\end{document}